\documentclass[twoside,11pt]{10ssmmp} 
\usepackage{fancyhdr}
\usepackage{amsfonts}
\usepackage{amssymb}
\usepackage{graphicx}
\usepackage{epsfig}

\newcommand\rf[1]{(\ref{eq:#1})}
\newcommand\lab[1]{\label{eq:#1}}
\newcommand\nonu{\nonumber}
\newcommand\br{\begin{eqnarray}}
\newcommand\er{\end{eqnarray}}
\newcommand\be{\begin{equation}}
\newcommand\ee{\end{equation}}

\newcommand\lb{\lbrack}
\newcommand\rb{\rbrack}

\renewcommand\){\right)}
\newcommand\bgv{\bigg\vert}              

\newcommand\bc{\begin{center}}
\newcommand\ec{\end{center}}

\newcommand\partder[2]{\frac{{\partial {#1}}}{{\partial {#2}}}}

\renewcommand\d{\delta}

\newcommand\vareps{\varepsilon}

\newcommand\G{\Gamma}

\newcommand\h{\frac{1}{2}}
\renewcommand\k{\kappa}
\renewcommand\l{\lambda}
\renewcommand\L{\Lambda}
\newcommand\m{\mu}
\newcommand\n{\nu}
\newcommand\om{\omega}

\renewcommand\P{\Phi}
\newcommand\pa{\partial}

\newcommand\pr{\prime}

\newcommand\z{\zeta}

\newcommand\twomat[4]{\left(\begD{array}{cc}  
{#1} & {#2} \\ {#3} & {#4} \end{array} \right)}

\newcommand\cM{{\mathcal M}}
\newcommand\cN{{\mathcal N}}

\newcommand{\ct}[1]{\cite{#1}}
\newcommand\PRep[3]{{Phys. Reports} \textbf{#1}, #3 (#2)}

\newcommand\PRL[3]{\textsl{Phys. Rev. Lett.} \textbf{#1} (#2) #3}

\newcommand\PRD[3]{\textsl{Phys. Rev.} \textbf{D#1} (#2) #3}

\newcommand\PLB[3]{\textsl{Phys. Lett.} \textbf{#1B} (#2) #3}
\newcommand\CQG[3]{\textsl{Class. Quantum Grav.} \textbf{#1} (#2) #3}

\newcommand\IJMPA[3]{\textsl{Int. J. Mod. Phys.} \textbf{A#1} (#2) #3}
\newcommand\IJMPD[3]{\textsl{Int. J. Mod. Phys.} \textbf{D#1} (#2) #3}

\newcommand\udot{\stackrel{.}{u}}
\newcommand\uddot{\stackrel{..}{u}}

\newcommand\adot{\stackrel{.}{a}}

\newcommand\Hdot{\stackrel{.}{H}}

\begin{document}


\title{Non-Riemannian Volume Elements Dynamically Generate Inflation}



\author{
\bf{David Benisty}\hspace{.25mm}\thanks{\,e-mail address: 
benidav@post.bgu.ac.il}
\\ \normalsize{Physics Department, Ben Gurion University of the Negev}\\
\normalsize{Beer Sheva, Israel} \vspace{2mm} \\
\normalsize{Frankfurt Institute for Advanced Studies (FIAS)}\\ 
\normalsize{Frankfurt am Main, Germany} \\
\bf{Eduardo Guendelman}\hspace{.25mm}\thanks{\,e-mail address: 
guendel@bgu.ac.il, guendelman@fias.uni-frankfurt.de}
\\ \normalsize{Physics Department, Ben Gurion University of the Negev}\\
\normalsize{Beer Sheva, Israel} \\
\normalsize{Frankfurt Institute for Advanced Studies (FIAS)}\\ 
\normalsize{Frankfurt am Main, Germany} \\
\normalsize{Bahamas Advanced Study Institute and Conferences} \\
\normalsize{Stella Maris, Long Island, The Bahamas} \vspace{2mm} \\
\bf{Emil Nissimov and Svetlana Pacheva}\hspace{.25mm}\thanks{\,e-mail
address: nissimov@inrne.bas.bg, svetlana@inrne.bas.bg}
\\ \normalsize{Institute for Nuclear Research and Nuclear Energy}\\
\normalsize{Bulgarian Academy of Sciences, Sofia, Bulgaria} }

\date{} 

\maketitle 

\begin{abstract}
Our primary objective is the formulation of a plausible cosmological inflationary
model entirely in terms of a pure modified gravity without any {\em a priori} 
matter couplings within the formalism of non-Riemannian spacetime volume elements. 
The non-Riemannian volume elements {\em dynamically} create 
in the physical Einstein frame a canonical scalar matter field and produce 
a non-trivial inflationary scalar potential with a large flat region and 
a low-lying stable minimum corresponding to the late universe stage. This
dynamically generated inflationary potential is a substantial generalization
of the classic Starobinsky potential. Our  model predicts scalar power spectral 
index and tensor to scalar ratio in accordance with the available
observational data.

\end{abstract}

\section{Introduction}

The theoretical framework based on the concept of ``inflation'' in the study
of the evolution of the early Universe provides an attractive solution 
explaining the ``puzzles'' of Big-Bang cosmology (the horizon problem, 
the flatness problem, the magnetic monople problem, etc. 
\ct{Starobinsky:1979ty}-\ct{Albrecht:1982wi}.
Likewise it is an important instrumentarium for treatment of primordial density 
perturbations \ct{Mukhanov:1981xt,Guth:1982ec}.
For some recent detailed accounts, see 
Refs.\ct{weinberg-2008}-\ct{mukhanov-winitzki}.

On the other hand, in a parallel development another groundbreaking concept
emerged in the last decade or so about the intrinsic necessity to modify
(extend) gravity theories beyond the scope of standard Einstein's general
relativity with the main motivation to overcome the limitations of the
latter coming from: (i) Cosmology -- for solving the problems of dark energy and 
dark matter and explaining the large scale structure of the Universe 
\ct{Perlmutter:1998np,Copeland:2006wr};
(ii) Quantum field theory in curved spacetime -- due to renormalization of
ultraviolet divergences in higher loops \ct{weinberg-79}; 
(iii) Modern string theory --
due to the natural appearance of higher-order curvature invariants and
scalar-tensor couplings in low-energy effective field theories \ct{GSW-1}.

Various classes of modified gravity theories have been employed to construct
viable inflationary models: $f(R)$-gravity; scalar-tensor gravity; 
Gauss-Bonnet gravity (see Refs.\ct{extended-grav-book}-\ct{odintsov-2} 
for a detailed accounts); recently also based on non-local gravity 
(Ref.\ct{dragovich-etal} and references therein) or based on brane-world
scenarios (Ref.\ct{djordjevich-etal} and references therein). 
The first early successful cosmological model based
on the extended $f(R)= R + R^2$-gravity is the classical Starobinsky
potential \ct{Starobinsky:1979ty}.

A further specific broad class of actively developed modified (extended)
gravitational theories is based on the formalism of 
{\em non-Riemannian spacetime volume elements} (originally proposed in
Refs.\ct{TMT-orig-0}-\ct{5thforce}; see 
Refs.\ct{susyssb-1,grav-bags} for a systematic geometric formulation). 
This formalism was used as a basis for constructing a series of 
extended gravity-matter
models describing unified dark energy and dark matter scenario \ct{dusty,dusty-2},
quintessential cosmological models with gravity-assisted and inflaton-assisted
dynamical suppression (in the ``early'' universe) or dynamical generation (in the
post-inflationary universe) of electroweak spontaneous symmetry
breaking and charge confinement \ct{grf-essay}-\ct{bpu-10}, as well as a novel 
mechanism for dynamical supersymmetric Brout-Englert-Higgs effect in supergravity 
\ct{susyssb-1}.

In what follows we will describe in some detail the construction of a viable
cosmological inflationary model starting from a modified pure gravity
involving several independent non-Riemannian volume elements and 
{\em without any a priori coupling} to matter fields.

\section{Brief Reminder on Non-Riemannian Volume-Forms (Volume Elements)}

Let us briefly recall the essence of the non-Riemannian volume-form formalism
(cf. Ref.\ct{spivak}).

In integrals over differentiable manifolds (not necessarily 
Riemannian one, so {\em no} metric is needed) volume-forms are given by 
nonsingular maximal rank differential forms $\om$:
\be
\int_{\cM} \om \bigl(\ldots\bigr) = \int_{\cM} dx^D\, \Omega \bigl(\ldots\bigr)
\;\;,\;\; 
\om = \frac{1}{D!}\om_{\m_1 \ldots \m_D} dx^{\m_1}\wedge \ldots \wedge dx^{\m_D}
\; ,
\lab{omega-1}
\ee
where  $\om_{\m_1 \ldots \m_D} = - \vareps_{\m_1 \ldots \m_D} \Omega$ and
$\Omega$ is the volume element density (our conventions for the totally
anti-symmetric symbols are $\vareps^{01\ldots D-1}=1 \; ,\;
\vareps_{01\ldots D-1}=-1$).

In Riemannian $D$-dimensional spacetime manifolds a standard generally-covariant 
volume-form is defined through the ``D-bein'' (frame-bundle) canonical one-forms 
$e^A = e^A_\m dx^\m$ ($A=0,\ldots ,D-1$):
\be
\om = e^0 \wedge \ldots \wedge e^{D-1} = \det\Vert e^A_\m \Vert\,
dx^{\m_1}\wedge \ldots \wedge dx^{\m_D} \; ,
\lab{omega-riemannian}
\ee
where the standard Riemannian volume element density reads
$\Omega = \det\Vert e^A_\m \Vert = \sqrt{-\det\Vert g_{\m\n}\Vert} \equiv \sqrt{-g}$.

To construct modified gravitational theories as alternatives to ordinary
standard theories in Einstein's general relativity, instead of $\sqrt{-g}$ 
we can employ one or more alternative {\em non-Riemannian} 
volume element(s) as in \rf{omega-1} given by non-singular {\em exact} $D$-forms
$\om = d A$,  where:
$A = \frac{1}{(D-1)!} A_{\m_1\ldots\m_{D-1}}
dx^{\m_1}\wedge\ldots\wedge dx^{\m_{-1}}$ and the corresponding
volume element density reads:
\be
\Omega \equiv \Phi(A) =
\frac{1}{(D-1)!}\vareps^{\m_1\ldots\m_D}\, \pa_{\m_1} A_{\m_2\ldots\m_D} \; .
\lab{Phi-D}
\ee
Thus, non-Riemannian volume element densities $\P (A)$ are defined in terms of
the (scalar density of the) dual field-strength of auxiliary rank 
$D-1$ tensor gauge fields $A_{\m_1\ldots\m_{D-1}}$. 

As an important remark, let us note that in the first-order (Palatini) formalism
($g_{\m\n}$ and $\G_{\m\n}^\l$ {\em a priori} independent), the auxiliary
tensor gauge fields $A_{\m_1\ldots\m_{D-1}}$ turn out to be (almost)
pure-gauge -- {\em no propagating field degrees of freedom} except for few
discrete degrees of freedom with conserved canonical momenta appearing as 
arbitrary integration constants. See Refs.\ct{grav-bags}-\ct{grf-essay} (appendices A)
for a systematic proof of the latter fact using the standard canonical 
Hamiltonian treatment of systems with gauge symmetries, \textsl{i.e.},
systems with first-class Hamiltonian constraints a'la Dirac
\ct{henneaux-teitelboim,rothe}.

However, in the second-order (metric) formalism 
(where $\G_{\m\n}^\l$ is the usual Levi-Civita connection of the metric $g_{\m\n}$)
the first non-Riemannian volume form $\P(A)$, replacing $\sqrt{-g}$ in the
modified Einstein-Hilbert part of the action:
\be
S= \int d^4 x \P(A) R + \ldots \; ,
\lab{E-H-modified}
\ee
is {\em not} any more pure-gauge. The reason is that in the action \rf{E-H-modified} 
the scalar curvature $R$ (in the metric formalism) containes {\em second-order} 
(time) derivatives (the latter amount to a total derivative in the ordinary case 
$S= \int d^4 x \sqrt{-g} R + \ldots$). 

So defining $\chi_1 \equiv \P(A)/\sqrt{-g}$, the latter field  
becomes physical degree of freedom as seen from the equations of motion
resulting from  varying \rf{E-H-modified} w.r.t. $g^{\m\n}$:
\be
R_{\m\n} + \frac{1}{\chi_1} \bigl(g_{\m\n}\Box\chi_1 
-\nabla_\m \nabla_\n \chi_1 \bigr) + \ldots = 0
\lab{chi1-degree-freedom}
\ee

\section{Modified Pure Gravity with Non-Riemannian Volume Elements}

Let us now consider the following simple modified gravity model without any
couplings to  matter fields (we will use ``Planck units'' $16\pi G_N = 1$):
\be
S = \int d^4 x \Bigl\{\P_1 (A)\Bigl\lb R 
- 2\L_0 \frac{\P_1 (A)}{\sqrt{-g}}\Bigr\rb 
+ \P_2 (B) \frac{\P_0 (C)}{\sqrt{-g}}\Bigr\} \; . 
\lab{NRVF-1}
\ee
Here $R$ is the scalar curvature in the metric formalism and:
\br
\P_1 (A) \equiv \frac{1}{3!}\vareps^{\m\n\k\l} \pa_\m A_{\n\k\l} \;,\;
\P_2 (B) \equiv \frac{1}{3!}\vareps^{\m\n\k\l} \pa_\m B_{\n\k\l} \;,
\nonu \\
\P_0 (C) \equiv \frac{1}{3!}\vareps^{\m\n\k\l} \pa_\m C_{\n\k\l} \; ,
\phantom{aaaaaa}
\lab{Phi-D4}
\er
denote three different independent non-Riemannian volume element denisties
as in \rf{Phi-D} for $D=4$.
$\L_0$ is dimensionful parameter which will turn out in what follows to play 
the role of an inflationary scale.

It is important to stress that the form of the action \rf{NRVF-1} is uniquely 
specified by the requirement about global Weyl-scale invariance under:
\br
g_{\m\n} \to \l g_{\m\n} \;, \; 
A_{\m\n\k} \to \l A_{\m\n\k} \; ,\; B_{\m\n\k} \to \l^2 B_{\m\n\k} \; ,\; 
C_{\m\n\k} \to C_{\m\n\k} \; .
\lab{scale-transf}
\er
where $\lambda = {\rm const}$. Its importance within the context of
non-Riemannian volume element formalism has been originally stressed in 
\ct{TMT-orig-1}.

The equations of motion from the action \rf{NRVF-1} w.r.t. the auxiliary gauge fields 
$A_{\m\n\l},\, B_{\m\n\l}\,, C_{\m\n\l}$ defining the non-Riemannian volume
elements \rf{Phi-D4} yield, respectively:
\br
R 
- 4\L_0 \frac{\P_1 (A)}{\sqrt{-g}} = - M_1 \equiv {\rm const} \; ,
\lab{A-eq} \\
\frac{\P_0 (C)}{\sqrt{-g}} = - M_2 \equiv {\rm const} \;\; , \;\; 
\frac{\P_2 (B)}{\sqrt{-g}} = \chi_2 \equiv {\rm const} \; .
\lab{C-B-eq}
\er
Here $M_1 , M_2$ and  $\chi_2$ are (dimensionful and dimensionless, respectively) 
free integration constants; $M_1 , M_2$ indicate
spontaneous breaking of global Weyl symmetry \rf{scale-transf}.

Also, it is important to observe that, since the scalar curvature $R$
contains terms with second-order time derivatives on $g_{\m\n}$, 
Eq.\rf{A-eq} is a genuine dynamical equation of motion and {\em not} a
constraint.

The equations of  motion w.r.t. $g_{\m\n}$ from \rf{NRVF-1} read:
\be
R_{\m\n} - \L_0\chi_1\, g_{\m\n}
+ \frac{1}{\chi_1}\bigl( g_{\m\n} \Box{\chi_1} - \nabla_\m \nabla_\n \chi_1\Bigr)
- \frac{\chi_2 M_2}{\chi_1} g_{\m\n} = 0 \; ,
\lab{einstein-like}
\ee
with $\chi_1 \equiv \P(A)/\sqrt{-g}$. Taking the trace of \rf{einstein-like}:
\be
3 \frac{\Box \chi_1}{\chi_1} - \frac{4\chi_2 M_2}{\chi_1} - M_1 = 0 
\lab{chi1-eq}
\ee
yields a dynamical equation of motion for the composite scalar field $\chi_1$.

\section{From Modified Gravity to the Physical Einstein Frame}

We now transform Eqs.\rf{einstein-like} and \rf{chi1-eq} to the physical
Einstein frame  via the conformal transformation 
${\bar g}_{\m\n} = \chi_1 g_{\m\n}$, upon using the well-known 
(cf. Ref.\ct{dabrowski}) conformal transformation formulas (bars indicate 
magnitudes in the ${\bar g}_{\m\n}$-frame):
\br
R_{\m\n}(g) = R_{\m\n}(\bar{g}) - 3 \frac{{\bar g}_{\m\n}}{\chi_1}
{\bar g}^{\k\l} \pa_\k \chi_1^{1/2} \pa_\l \chi_1^{1/2} 
\nonu \\
+ \chi_1^{-1/2}\bigl({\bar \nabla}_\m {\bar \nabla}_\n \chi_1^{1/2} +
{\bar g}_{\m\n} {\bar{\Box}}\chi_1^{1/2}\bigr) \; ,
\lab{dabrowski-1} \\
\Box \chi_1 = \chi_1 \Bigl({\bar{\Box}}\chi_1 
- 2{\bar g}^{\m\n} \frac{\pa_\m \chi_1^{1/2} \pa_\n \chi_1}{\chi_1^{1/2}}\Bigr)
\; .
\lab{dabrowski-2} 
\er
Hereby the  transformed 
equations acquire the standard form of Einstein equations w.r.t. the 
new ``Einstein-frame'' metric ${\bar g}_{\m\n}$:
\br
R_{\m\n}(\bar{g}) - \h {\bar g}_{\m\n} R(\bar{g}) =
\h \Bigl\lb \pa_\m u \pa_\n u
- {\bar g}_{\m\n}\bigl(\h {\bar g}^{\k\l} \pa_\k u \pa_\l u
+ U_{\rm eff} (u)\bigr)\Bigr\rb \; ,
\lab{EF-eqs} \\
{\bar{\Box}} u + \partder{U_{\rm eff}}{u} = 0 \; ,
\lab{u-eq-orig}
\er
where we have redefined
\be
\P_1 (A)/\sqrt{-g}\equiv \chi_1 = \exp{\bigl(u/\sqrt{3}\bigr)}
\lab{u-def}
\ee
in order to obtain a canonically normalized kinetic term for the scalar
field $u$, and where we have obtained a {\em dynamically generated effective scalar 
potential}:
\be
U_{\rm eff} (u) = 2 \L_0 - M_1 \exp{\bigl(-\frac{u}{\sqrt{3}}\bigr)} 
+ \chi_2 M_2 \exp{\bigl(-2 \frac{u}{\sqrt{3}}\bigr)} \; .
\lab{U-eff}
\ee
$U_{\rm eff}$ \rf{U-eff} is a generalization of the classic {\em Starobinsky
potential} \ct{Starobinsky:1979ty}; the latter is a special case of \rf{U-eff} for
$\L_0 = \frac{1}{4}M_1 = \h \chi_2 M_2$.

Accordingly, the corresponding Einstein-frame action reads:
\be
S_{\rm EF} = \int d^4 x \sqrt{-{\bar g}} \Bigl\lb R({\bar g}) 
- \h {\bar g}^{\m\n}\pa_\m u \pa_\n u - U_{\rm eff} (u) \Bigr\rb\; ,
\lab{EF-action}
\ee
with $U_{\rm eff}$ as in \rf{U-eff}.

Let us particularly emphasize that the Einstein-frame action \rf{EF-action}
is entirely dynamically generated:

$\phantom{aaa}$(a) The canonical scalar field $u$ is dynamically created from the 
ratio of the volume-element densities $\P_1(A)/\sqrt{-g}$ \rf{u-def};

$\phantom{aaa}$(b) The effective potential $U_{\rm eff}(u)$ \rf{U-eff} is 
dynamically generated due to the appearance of the free integration constants 
$M_{1,2}, \chi_2$ in \rf{U-eff} as a result of the specific (constrained)
dynamics \rf{A-eq}-\rf{C-B-eq} of the auxiliary gauge fields
$A_{\m\n\l}, B_{\m\n\l}, C_{\m\n\l}$ -- constituents of the non-Riemannian 
volume element densities $\P (A), \P (B), \P (C)$ \rf{Phi-D4}. 
$U_{\rm eff}(u)$ \rf{U-eff} is graphically depicted on Fig.1.

\begin{figure}
\begin{center}
\includegraphics[width=9cm,keepaspectratio=true]{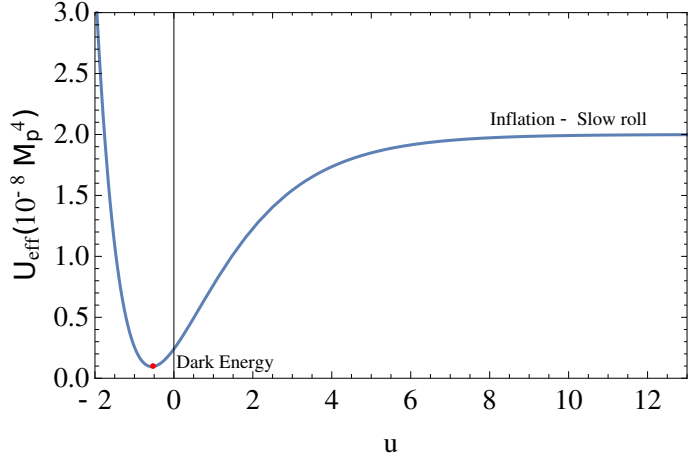}
\caption{Qualitative shape of the dynamically generated effective scalar potential 
$U_{\rm eff}$ \rf{U-eff} as function of $u$. 
The unit for $u$ is $M_{Planck}/\sqrt{2}$.}
\end{center}
\label{fig1}
\end{figure}

The dynamically generated potential $U_{\rm eff}(u)$ \rf{U-eff} has two main 
features relevant for cosmological applications.

First, $U_{\rm eff} (u)$ \rf{U-eff} possesses a long flat region
for large positive $u$ and, second, it has a stable minimum for a small 
finite value $u=u_{*}$:
\begin{itemize}
\item
(i) $U_{\rm eff} (u) \simeq 2\L_0$ for large $u$;
\item
(ii) $\partder{U_{\rm eff}}{u} = 0\;$ for $u\equiv u_{*}$ where:
\be
\exp\bigl(-\frac{u_{*}}{\sqrt{3}}\bigr) = \frac{M_1}{2\chi_2 M_2}  \quad ,\quad 
\frac{\pa^2 U_{\rm eff}}{\pa u^2}\bgv_{u=u_{*}} = \frac{M_1^2}{6\chi_2 M_2} >0 \; . 
\lab{stable-min}
\ee
\end{itemize}

The flat region of $U_{\rm eff} (u)$ for large positive $u$ correspond to
``early'' universe' slow-roll inflationary evolution with energy scale $2\L_0$. 
On the other hand, the region around the stable minimum at $u=u_{*}$ \rf{stable-min}
correspond to ``late'' universe' evolution where: 
\be
U_{\rm eff} (u_{*})= 2\L_0 - \frac{M_1^2}{4\chi_2 M_2} \equiv 2 \L_{\rm DE}
\lab{DE-value}
\ee
is the dark energy density value dynamically generated through the free 
integration constants $M_{1,2}, \,\chi_2$.

\section{FLRW Reduction and Evolution of the Homogeneous Solution}

Let us mow consider the reduction of the Einstein-frame action \rf{EF-action} to the
Friedmann-Lemaitre-Robertson-Walker (FLRW) setting with metric
$ds^2 = - N^2 dt^2 + a(t)^2 d{\vec x}^2$, and with $u=u(t)$.

The FLRW-reduced action reads:
\be
S_{\rm FLRW} = \int d^4 x \Bigl\lb - 6\frac{a\adot^2}{N}
+ N a^3 \Bigl(
\h \udot^2 
+ M_1 e^{-u/\sqrt{3}} - M_2\chi_2 e^{-2u/\sqrt{3}} - 2\L_0 \Bigr)\Bigr\rb \; .
\lab{EF-action-FLRW}
\ee
We will study the evolution of $u=u(t)$ and $a=a(t)$ specified by 
\rf{EF-action-FLRW} using the method of autonomous dynamical systems. 

The pertinent Friedmann and $u$-field equations resulting from \rf{EF-action-FLRW}
are given by:
\br
H^2 = \frac{1}{6}\rho \;\; ,\;\; \rho = 
\h\udot^2 + U_{\rm eff}(u)\; , 
\lab{Fried-1} \\
\Hdot = - \frac{1}{4} (\rho + p) 
\;\; ,\;\; p = 
\h\udot^2 - U_{\rm eff}(u)\; ,
\lab{Fried-2} \\
\uddot + 3H \udot + \partder{U_{\rm eff}}{u} = 0 \; .
\lab{u-eq}  
\er

It is instructive (following Ref.\ct{Bahamonde:2017ize}) to rewite the system of 
Eqs.\rf{Fried-1}-\rf{u-eq} in terms of a set of dimensionless variables:
\be
x := \frac{\dot{u}}{\sqrt{12} H},\quad 
y := \frac{\sqrt{U_{\rm eff}(u) - 2\L_{\rm DE}}}{\sqrt{6} H}, \quad 
z := \frac{\sqrt{\L_{\rm DE}}}{\sqrt{3}H} \; ,
\lab{xyz-def}
\ee
with $L_{\rm DE} =\L_0 - \frac{M_1^2}{8\chi_2 M_2}$ as in \rf{DE-value}.

The first Friedman Eq.\rf{Fried-1} yields an algebraic constraint
$x^2 + y^2 + z^2 = 1$, so that the autonomous dynamical system w.r.t. 
$(x,z)$ reads:
\br
x' = \frac{\sqrt{3}}{2 \Lambda_{DE}} z^2 \left[ -M_1 \xi(x,z)
+2M_2 \chi_2 \xi^2(x,z)\right] 
-3x (1-x^2) \; , 
\nonu \\
z' = 3zx^2 \quad ,\phantom{aaaaaaaaaaaa} 
\lab{asm}
\er
where the primes denote derivative w.r.t. the  parameter $\cN = \log a$
(number of $e$-foldings), and the function $\xi(x,z)$ is defined as: 
\be
\xi(x,z) = \frac{M_1}{2\chi_2 M_2} \Bigl\lb 1 -
\sqrt{\frac{8\L_0 M_2\chi_2}{M_1^2}\,\frac{1-x^2 -z^2}{z^2}}\Bigr\rb \; .
\lab{zeta-def}
\ee

Phase space portrait of the autonomous system \rf{asm} is depicted
numerically on Fig.2.

\begin{figure}
\begin{center}
\includegraphics[width=9cm,keepaspectratio=true]{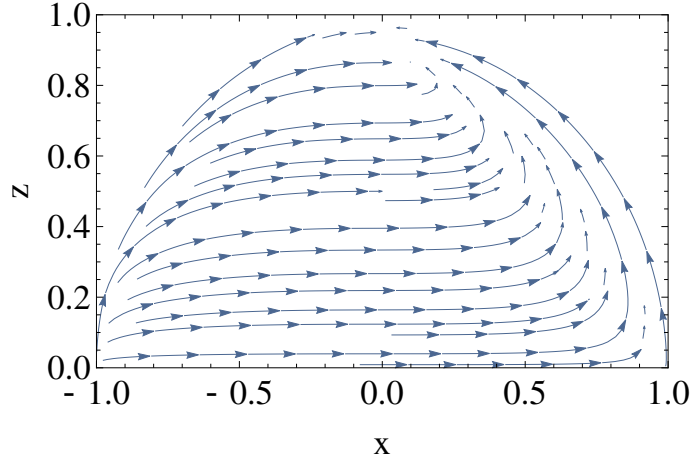}
\caption{Phase space portrait of the autonomous system \rf{asm}. 
The $x$ axis denotes the relative kinetic part of the scalar inflaton,
and the $z$ axis denotes the relative part of the dark energy density 
$\L_{\rm DE}$.}
\end{center}
\label{fig2}
\end{figure}

The autonomous system \rf{asm} possesses the following two
critical  points:

(a) Stable critical point $A\(x=0, z=1 \)$ corresponding to
the ``late'' universe de Sitter behavior with a cosmological constant
$\L_{\rm DE}$ \rf{DE-value}.

(b) Unstable critical point $B\(x=0,z=\sqrt{\L_{\rm DE}/\L_0}\)$
corresponding to beginning of  evolution in the ``early'' universe at
large $u$. If the evolution starts at any point close to $B$,
then the dynamics drives the system away from $B$ all the way 
towards the stable point $A$ at late times.

Numerical solutions of the FLRW system \rf{Fried-1}-\rf{u-eq}
are graphically presented on
Fig.3 for the Hubble parameter $H(t)$, and on Fig.4 for the scalar field $u(t)$.

\begin{figure}
\begin{center}
\includegraphics[width=9cm,keepaspectratio=true]{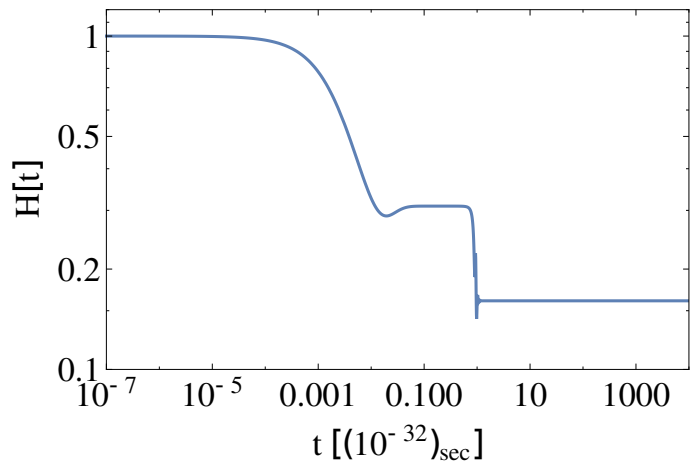}
\caption{Numerical example of the solution for the Hubble parameter $H(t)$ vs. time. 
Initially for short times the inflationary Hubble parameter is large and afterwards 
approaches its cosmological late time value.}
\end{center}
\label{fig3}
\end{figure}

\begin{figure}
\begin{center}
\includegraphics[width=9cm,keepaspectratio=true]{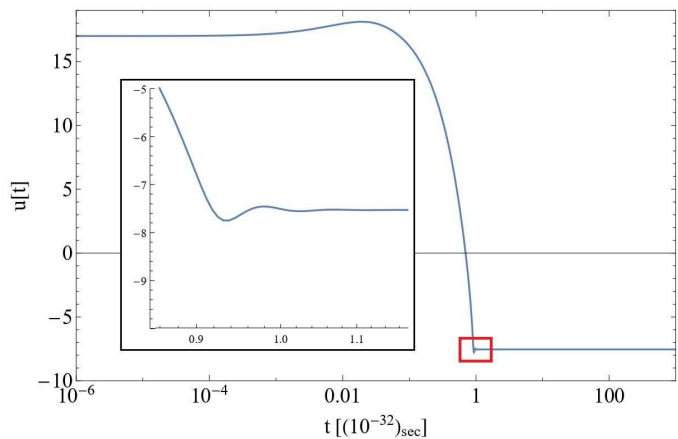}
\caption{Numerical example of the solution for the scalar field $u(t)$ vs. time.
The unit for $u$ is $M_{Planck}/\sqrt{2}$.
The blown-up rectangle depicts the oscillations of $u(t)$ around the minimum of
$U_{\rm eff}$ \rf{U-eff}.}
\end{center}
\label{fig4}
\end{figure}

\section{Perturbations and Observables}

In order to check the viability of our model we will investigate the perturbations of 
the above FLRW background evolution \rf{Fried-1}-\rf{u-eq}, in particular 
focusing on the inflationary observables such as the scalar power spectral 
index $n_s$ and the tensor-to-scalar ratio $r$ (for definitions, see
\textsl{e.g.} Ref.\ct{Nojiri:2019kkp}). 
As usual, we introduce the Hubble slow-roll parameters, which in our case using the potential $U_{\rm eff}(u)$ \rf{U-eff} read:
\br
\epsilon = \Bigl(\frac{U_{\rm eff}^\pr (u)}{U_{\rm eff}(u)}\Bigr)^2
= \frac{4\z^2}{3} \frac{\bigl(1/2 - \z\bigr)^2}{\bigl\lb \bigl(1/2 - \z\bigr)^2
+ \d/4\bigr\rb^2} \; ,
\lab{eps-1} \\
|\eta| = 2 |\frac{U_{\rm eff}^{\pr\pr}(u)}{U_{\rm eff}(u)}|
= \frac{2\z}{3} \frac{\bigl(1-4\z\bigr)}{\bigl\lb \bigl(1/2 - \z\bigr)^2
+ \d/4\bigr\rb} \; ,
\lab{eta-1}
\er
where:
\be
\z\equiv \frac{M_2 \chi_2}{M_1}\,e^{-u/\sqrt{3}} \quad,\quad
\d \equiv \frac{8M_2 \chi_2}{M_1^2} \L_{\rm DE} \; ,
\lab{zeta-delta-def}
\ee
with $\L_{\rm DE}$ -- the dark energy density \rf{DE-value}, and therefore the
parameter $\d$ being very small.

Inflation ends when $\epsilon (u_f) = 1$ for some $u=u_f$ whose value
(using the short-hand notation 
$\z_f \equiv \frac{M_2 \chi_2}{M_1} e^{-u_f/\sqrt{3}}$) is given by:
\be
\z_f =
\frac{1}{2\bigl(1+2/\sqrt{3}\bigr)} \Bigl\lb 1+\frac{1}{\sqrt{3}}
- \sqrt{1/3 - \bigl(1+2/\sqrt{3}\bigr)\d} \Bigr\rb 
\simeq \frac{1}{2\bigl(1+2/\sqrt{3}\bigr)} \; . 
\lab{z-f}
\ee

For the number of $e$-foldings 
$\cN = \h \int_{u_i}^{u_f} du \; U_{\rm eff}/ U_{\rm eff}^\pr$ we obtain:
\be
\cN = \frac{3}{8}(1+\d)\Bigl(1/\z_i - 1/\z_f\Bigr) 
- \frac{3}{4}(1-\d) \log\frac{\z_f}{\z_i}
+ \frac{3}{4}\d\,\log\Bigl(\frac{1-2\z_i}{1-2\z_f}\Bigr) \; ,
\lab{N-def}
\ee
where $\z_i \equiv \frac{M_2 \chi_2}{M_1} e^{-u_i/\sqrt{3}}$ and $u=u_i$ is
very large corresponding to the start of the inflation. 

Ignoring the very small $\d$ we have for $\cN$ approximately:
\be
\cN \simeq \frac{3M_1}{8M_2\chi_2} e^{u_i/\sqrt{3}} - \frac{\sqrt{3}}{4} u_i
- \frac{3}{4} \bigl(1+2/\sqrt{3}\bigr) 
+ \frac{3}{4} \log\Bigl(2\bigl(1+2/\sqrt{3}\bigr)\Bigr) \; .
\lab{N-approx}
\ee
Using the slow-roll parameters \rf{eps-1}-\rf{eta-1}, one can calculate the 
values of the scalar spectral index $n_s$  and the tensor-to-scalar ratio $r$, 
respectively, as functions of the $e$-foldings $\cN$:  
\be
r \approx 16\,\epsilon \bigl(u_i(\cN)\bigr) \quad, \quad 
n_s \approx 1 - 6\,\epsilon \bigl(u_i(\cN)\bigr) + 2 \eta\,\bigl(u_i(\cN)\bigr)
\; ,
\lab{r-ns-def}
\ee
where $u_i(\cN)$ is the solution of the transcedental Eq.\rf{N-approx} 
for $u_i$ as a function of $\cN$. From \rf{r-ns-def}, \rf{N-approx}, 
\rf{eps-1}, \rf{eta-1} we find:
\br
r \simeq \frac{12}{\Bigl\lb \cN + \frac{\sqrt{3}}{4} u_i(\cN) + c_0\Bigr\rb^2} 
\;\;\; ,\;\;
c_0 \equiv 
\frac{\sqrt{3}}{2} - \frac{3}{4} \log\Bigl(2\bigl(1+2/\sqrt{3}\bigr)\Bigr) \; ,
\lab{r-approx} \\
n_s \simeq 1 -\frac{r}{4}-\sqrt{\frac{r}{3}} \;\; . \phantom{aaaaaaaaaa}
\lab{n-s-approx}
\er

The numerical results for \rf{r-approx}-\rf{n-s-approx} are depicted on Fig.5.

\begin{figure}
\begin{center}
\includegraphics[width=9cm,keepaspectratio=true]{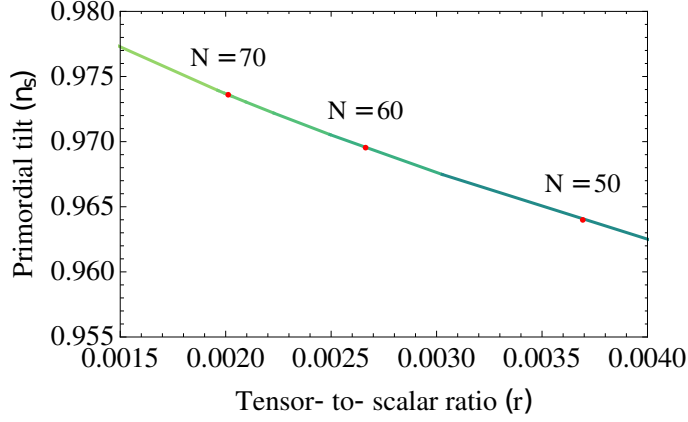}
\caption{The predicted values of the $r$ and $n_s$ for different number 
of $e$-foldings}
\end{center}
\label{fig5}
\end{figure}

The different values of the $r$ and $n_s $ are compatible with the 
PLANCK observational data ($0.95 < n_s < 0.97\; , \; r < 0.064$) 
(cf. Ref.\ct{Akrami:2018odb}).

Indeed, for the viable example of $\cN = 60$ $e$-foldings until the end of inflation
we obtain from \rf{N-approx}-\rf{n-s-approx}: 
\be
n_s \approx 0.969 \quad ,\quad r \approx 0.002 \; .
\lab{viable-values}
\ee

\section{Conclusions}

\begin{itemize}
\item
We proposed a very simple modified gravity model without any initial
coupling to  matter fields in terms of several alternative 
non-Riemannian spacetime volume elements within the second order (metric) 
formalism.
\item
We show how the non-Riemannian volume elements,
when passing to the physical Einstein frame, create a canonical scalar
field and produce dynamically a non-trivial inflationary-type potential for
the latter possesing a large flat region describing slow-roll inflation 
and a stable low-lying minimum corresponding to the late universe stage.
\item
We study the evolution and stability of the cosmological solutions from the point
of view of the theory of dynamical systems. Our model predicts scalar spectral index 
$n_s \approx 0.969$ and tensor-to-scalar ratio $r \approx 0.002$ for 60 $e$-folds, 
which is in accordance with the observational data.
\end{itemize}

\section*{Acknowledgments} 
E.N. and S.P. are sincerely grateful to Prof. Branko Dragovich, Prof. Marko Vojinovich
and all the organizers of the {\em Tenth Meeting in Modern Mathematical Physics} in 
Belgrade for cordial hospitality. 
We all  gratefully acknowledge support of our collaboration through 
the academic exchange agreement between the Ben-Gurion University in Beer-Sheva,
Israel, and the Bulgarian Academy of Sciences. 
D.B., E.N. and E.G. have received partial support from European COST actions
CA15117, CA16104 and CA18108. 
E.N. and S.P. are also thankful to Bulgarian National Science Fund for
support via research grant DN-18/1. 


\end{document}